\documentclass[twocolumn,showpacs,preprintnumbers,amsmath,prl,amssymb,superscriptaddress]{revtex4}

\usepackage{graphicx}
\usepackage{dcolumn}
\usepackage{bm}

\def\th232{\rm{ ^{232} Th }}
\def\u238{\rm{ ^{238} U }}
\def\cs137{\rm{^{137} Cs }}
\def\ba133{\rm{^{133} Ba }}
\def\cpd{\rm{kg^{-1}keV^{-1}day^{-1}}}
\def\nuebar{\rm{\bar{\nu_e}}}
\def\nue{\rm{\nu_e}}
\def\s2tw{\rm{ sin ^2 \theta _W }}

\def\munuebar{\rm{\mu_{\nuebar}}}
\def\mub{\rm{\mu_B}}
\def\enu{\rm{E_{\nu}}}

\def\ke10{\rm{\kappa_e}}

\begin{document}

\preprint{AS-TEXONO/02-07}

\title{
New Limits on Neutrino Magnetic Moments from\\
the Kuo-Sheng Reactor Neutrino Experiment
}

%
\newcommand{\as}{Institute of Physics, Academia Sinica, Taipei 115, Taiwan.} 
\newcommand{\ntu}{Department of Physics, National Taiwan University,
Taipei 106, Taiwan.}
\newcommand{\ihep}{Institute of High Energy Physics, Beijing 100039, China.}
\newcommand{\thu}{Department of Engineering Physics, Tsing Hua University,
Beijing 100084, China.}
\newcommand{\umd}{Department of Physics, University of Maryland,
College Park MD 20742, U.S.A.}
\newcommand{\ks}{Kuo-Sheng Nuclear Power Station, 
Taiwan Power Company, Kuo-Sheng 207, Taiwan.}
\newcommand{\iner}{Institute of Nuclear Energy Research, 
Lung-Tan 325, Taiwan.}
\newcommand{\ckit}{Department of Management Information Systems,
Chung Kuo Institute of Technology, Hsin-Chu 303, Taiwan.}
\newcommand{\ciae}{Department of Nuclear Physics, 
Institute of Atomic Energy, Beijing 102413, China.}
\newcommand{\corr}{htwong@phys.sinica.edu.tw}

\affiliation{ \as }
\affiliation{ \ntu }
\affiliation{ \ihep }
\affiliation{ \thu }
\affiliation{ \umd }
\affiliation{ \ks }
\affiliation{ \iner }
\affiliation{ \ckit }
\affiliation{ \ciae }

\author{ H.B. Li }  \affiliation{ \as } \affiliation{ \ntu }
\author{ J. Li }  \affiliation{ \as } \affiliation{ \ihep } \affiliation{ \thu }
\author{ H.T.Wong } \altaffiliation[Corresponding Author: ]{ \corr } \affiliation{ \as }
\author{ C.Y. Chang } \affiliation{ \as } \affiliation{ \umd }
\author{ C.P. Chen } \affiliation{ \as }
\author{ J.M. Fang } \affiliation{ \ks }
\author{ C.H. Hu } \affiliation{ \iner }
\author{ W.S. Kuo } \affiliation{ \iner }
\author{ W.P. Lai } \affiliation{ \as } \affiliation{ \ckit }
\author{ F.S. Lee } \affiliation{ \as }
\author{ S.C. Lee } \affiliation{ \as }
\author{ S.T. Lin } \affiliation{ \as }
\author{ C.S. Luo } \affiliation{ \as }
\author{ Y. Liu } \affiliation{ \ihep }
\author{ J.F. Qiu } \affiliation{ \as } \affiliation{ \ihep }
\author{ H.Y. Sheng } \affiliation{ \as } \affiliation{ \ihep }
\author{ V. Singh } \affiliation{ \as }
\author{ R.F. Su } \affiliation{ \ks }
\author{ P.K. Teng } \affiliation{ \as }
\author{ W.S. Tong } \affiliation{ \iner }
\author{ S.C. Wang } \affiliation{ \as }
\author{ B. Xin } \affiliation{ \as } \affiliation{ \ciae }
\author{ T.R. Yeh } \affiliation{ \iner }
\author{ Q. Yue } \affiliation{ \ihep }
\author{ Z.Y. Zhou } \affiliation{ \ciae }
\author{ B.A. Zhuang } \affiliation{ \as } \affiliation{ \ihep }

\collaboration{TEXONO Collaboration}

\noaffiliation


\date{\today}

\begin{abstract}
A laboratory has been set up 
at the Kuo-Sheng Nuclear Power Station
at a distance of 
28~m from the 2.9~GW reactor core 
to study low energy neutrino physics.
A detector threshold of 5~keV and a background level
of 1~$\cpd$ at 12-60~keV was achieved with a high purity germanium
detector of mass 1.06~kg surrounded by anti-Compton detectors with NaI(Tl)
and CsI(Tl) crystal scintillators.
Using 4712 and 1250 hours of Reactor ON and OFF data, respectively,
limits of the neutrino magnetic moment of
$\rm{ \munuebar < 1.3(1.0) \times 10^{-10} ~ \mub}$
at 90(68)\%  confidence level were derived.
Indirect bounds of the $\nuebar$ radiative lifetime
of $\rm{m_{\nu}^3 ~ \tau_{\nu} > 2.8(4.8) \times 10^{18} ~ eV^3 s}$
can be inferred.
\end{abstract}

\pacs{14.60.Lm, 13.15.+g, 13.40.Em}

\maketitle

The strong evidence of neutrino oscillations
from the solar and atmospheric neutrino measurements
implies finite neutrino masses and mixings~\cite{pdg,nu02}.
Their physical origin and experimental consequences
are not fully understood.
Experimental studies on the neutrino properties
and interactions can shed light to these
fundamental questions and/or constrain theoretical models.
The coupling of neutrinos with the photons are consequences of
non-zero neutrino masses.
Two of the manifestations
of the finite electromagnetic form factors~\cite{vogelengel}
are neutrino magnetic moments and radiative decays.
In this paper, we report new limits on these parameters
from data of a reactor neutrino experiment
taken at a lower energy range than previous measurements.

The searches of neutrino magnetic moments 
are performed in experiments on
neutrino-electron scatterings~\cite{kaiser}:
$\rm{
 \nu_{\it l_1}   +    e^-   \rightarrow   \nu_{\it l_2}   +   e^-  .
  }$
Both {\it diagonal} and {\it transition}
moments are allowed, corresponding to the cases where
$l_1=l_2$ and $l_1 \neq l_2$, respectively.
The experimental observable is the kinetic energy of the
recoil electrons (T).
The differential cross section 
for the magnetic scattering (MS) channel
is given by~\cite{vogelengel}:
\begin{equation}
\label{eq::mm}
\rm{
( \frac{ d \sigma }{ dT } ) _{MS}  ~ = ~
\frac{ \pi \alpha _{em} ^2 {\it \mu_l} ^2 }{ m_e^2 }
 [ \frac{ 1 - T/E_{\nu} }{T} ] ~ 
}
\end{equation}
where $\enu$ is the neutrino energy.
The neutrino magnetic moment ($\mu_l$),
often expressed
in units of the Bohr magneton($\mub$),
has an 1/T dependence and dominates
at low electron recoil energy over 
the Standard Model (SM) process.
The quantity $\mu_l$ is an effective parameter which,
in the case of $\nuebar$, 
can be expressed as~\cite{beacom}:
$\rm{ 
\munuebar ^2  =  \sum_j \big| \sum_k U_{ek}  \mu_{jk} \big| ^2 
}$
where U is the mixing matrix and $\rm{\mu_{jk}}$
are the fundamental constants that characterize
the couplings between the mass eigenstates 
$\rm{\nu_j}$ and $\rm{\nu_k}$ with the
electromagnetic field.
In this paper, we write
$\rm{\munuebar = \ke10 \times 10^{-10} ~ \mub}$
for simplicity.
The neutrino-photon couplings
probed by $\nu$-e scatterings
are related to those giving rise to the
neutrino radiative decays~\cite{rdk}:
$\rm{
\nu_j  \rightarrow   \nu_k   +   \gamma    
}$
between $\rm{\nu_j}$ and $\rm{\nu_k}$.
The decay rate $\rm{\Gamma_{jk}}$ is
given by by:
$
\rm{
\Gamma_{jk} =  
[ \mu_{jk}^2  ~ ( m_j^2 - m_k^2 ) ^ 3 ] /
[ 4 \pi ~ m_j^3 ]  
}
$,
where $\rm{m_i}$ is the mass of $\rm{\nu_i}$.

Reactor neutrinos provide a sensitive probe
for ``laboratory'' searches  of $\rm{\munuebar}$,
taking advantages of the
high $\nuebar$ flux, low $\enu$ and the better experimental
control via the reactor ON/OFF comparison.
A finite $\rm{\munuebar}$ would manifest itself
as excess events in the ON over OFF periods,
where the energy spectra have an 1/T profile.
Neutrino-electron scatterings were first
observed in the pioneering experiment~\cite{reines}
at Savannah River.  
A revised analysis of the data by Ref~\cite{vogelengel}
with improved input parameters
gave a positive signature
consistent with the interpretation of a
finite $\munuebar$ at $\ke10$=2$-$4.
Other results came
from the Kurtchatov~\cite{kurt} and Rovno~\cite{rovno} experiments
which quoted limits of $\ke10 < 2.4$
and $< 1.9$ at 90\% confidence level (CL), respectively.
However, many experimental details, in particular
the effects due to the uncertainties in the SM ``background'',
were not discussed. 
An on-going experiment MUNU~\cite{munu}
is performed
at the Bugey reactor using
a time projection chamber with CF$_4$ gas.
A preliminary result of $\ke10 < 1.3$ (90\% CL)
was derived, while excess of events  below 1~MeV
during reactor ON was also reported.

Neutrino flavor conversion induced by
resonant or non-resonant spin-flip transitions
in the Sun via its transition magnetic moments
has been considered to
explain the solar neutrino measurements~\cite{vogelengel},
and were recently shown~\cite{sfpsnunew} to be
consistent with the existing data.
Alternatively, the measured solar
neutrino $\nu_{\odot}$-e spectral shape 
has been used to set limit of $\rm{\kappa_{\odot} < 1.5}$
at 90\% CL for the  ``effective'' $\nu_\odot$
magnetic moment~\cite{beacom} 
which is in general different
from that of a pure $\nue$ state derived in reactor experiments.
Astrophysical arguments~\cite{pdg,vogelengel},
placed bounds of $\rm{\kappa_{astro} < 10^{-2} - 10^{-3}}$,
but there are model dependence and implicit assumptions
on the neutrino properties involved.

Further laboratory experiments to put the
current limits on more solid grounds and to
improve on the sensitivities are therefore necessary.
The Kuo-Sheng (KS) results
described in this article
are based on data taken with
an ultra-low-background high purity germanium
detector (HPGe)
during the period June 2001 till April 2002.
The HPGe is optimal 
for magnetic moment searches~\cite{texono,munuruss}
with its low detection threshold,
excellent energy resolution, and robust stability.
The strategy is to focus on energy 
$<$100~keV~\cite{lernu}
where  the SM background are negligible
at the $10^{-10} ~ \mub$ range considered here.
A CsI(Tl) scintillating crystal array of mass 46~kg was
operating in parallel, and data were taken 
by a single readout system.

The KS neutrino laboratory~\cite{texono} is located at a distance
of 28~m from the Core \#1 of the Kuo-Sheng Power
Station in Taiwan. The nominal thermal power output
is 2.9~GW. The laboratory
is at the ground floor of the reactor building
at a depth of 12~m below sea-level and with
about 25~m water equivalence of overburden. 
The primary cosmic ray hadronic components
are eliminated while the muon flux is reduced
by a factor of 4. The laboratory is equipped
with an outer 50-ton shielding structure,
consisting of,
from outside in,
2.5~cm thick plastic scintillator panels with
photo-multipliers (PMTs) readout
for cosmic-ray veto (CRV),
15~cm of lead, 5~cm of stainless
steel support structures, 25~cm of boron-loaded polyethylene
and 5~cm of OFHC copper.
The innermost space has a dimension of
100$\times$80$\times$75~$\rm{cm^3}$
where both the HPGe and CsI(Tl) detectors and
their inner shieldings were placed.
Ambient $\gamma$-background at the reactor site
is about 20 times higher than that of a typical laboratory,
predominantly from radioactive dust with
$^{60}$Co and $^{54}$Mn.
The dust can get settled on exposed surfaces 
within hours and is difficult to remove.
Some earlier prototype detectors were contaminated
by such. Attempts to clean the surfaces on site
resulted in higher contaminations.
Accordingly, the various detector and
inner shielding components
were wrapped by plastic sheets during transportation,
to be removed on site only prior to installation.
Neutron background is the same as that of a typical
surface site, with no observable difference between the
ON and OFF period.

\begin{figure}
\includegraphics[width=7cm]{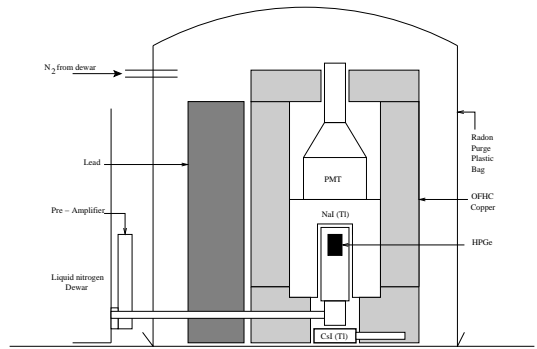}
\caption{
Schematic layout of the HPGe
with its anti-Compton detectors
as well as inner shieldings and
radon purge system.
}
\label{hpge}
\end{figure}

The HPGe set-up is schematically shown in
Figure~\ref{hpge}.
It is a coaxial germanium detector with an active
target mass of 1.06~kg. The lithium-diffused outer
electrode is 0.7~mm thick. The end-cap cryostat, also
0.7~mm thick, is made of OFHC copper. Both of these
features provide total suppression to ambient $\gamma$-background
below 60~keV, such that events below this energy are either
due to internal activity or ambient MeV-range $\gamma$'s 
via Compton scattering. 
The HPGe was surrounded by an anti-Compton (AC) detector system
made up of two components: (1) a NaI(Tl) well-detector
of thickness 5~cm that fit onto the end-cap cryostat,
the inner wall of which is also made of OFHC copper,
and (2) a 4~cm thick CsI(Tl) detector at the bottom.
Both AC detectors were read out by PMTs
with low-activity glass.
The assembly was surrounded by 3.7~cm of OFHC copper inner
shielding. Another 10~cm of lead provided additional
shieldings on the side of the liquid nitrogen dewar and
pre-amplifier electronics.
The inner shieldings and detectors were covered by a plastic
bag connected to the exhaust line of the dewar, serving
as a purge for the radioactive radon gas.

The electronics and data acquisition (DAQ) systems
of KS laboratory have been described elsewhere~\cite{eledaq}.
The HPGe pre-amplifier signals were distributed to
two spectroscopy amplifiers 
at the same 4~$\mu$s shaping time but with different
gain factors.
A low threshold on the output
provided the on-line trigger, ensuring all the events
down to the electronics noise edge of 5~keV were recorded.
The amplifier and the PMT signals
from the AC detectors  were recorded by 20~MHz Flash
Analog to Digital Convertor (FADC) modules
for a duration of 10~$\mu$s and 25~$\mu$s before and 
after the trigger, respectively.
The discriminator output of the CRV PMTs were
also recorded.
A random trigger  was provided by an external
clock at 0.1~Hz for sampling the pedestals
and for accurate measurements of
the efficiency factors.
The DAQ system remained active for 2~ms after
a trigger to  record possible time-correlated signatures.
The typical data taking rate for the HPGe sub-system
was about 1~Hz. The DAQ dead time was about 
10-20~ms per event and the typical system
live time was 96\%. 

Scatterings of $\nuebar$-e inside the Ge target
would manifest as ``lone-events'' uncorrelated 
with other detector systems.
The selection procedures, suppression
factors and efficiencies are summarized in Table~\ref{tabselect}.
The AC and CRV cuts suppressed Compton scattering
and cosmic-ray induced events.
The pulse shape analysis  
identified background due to electronic noise 
as well as the delayed ``cascade'' events.
The survival probabilities of the
random events along the various stages of the 
analysis procedures provided accurate measurements
of the DAQ dead time and the analysis efficiencies,
and allowed corrections to the residual instabilities of
the detector hardware.
The various residual $\gamma$-lines 
such as that of $^{40}$K provided the 
energy calibration as well as
independent consistency checks to
the efficiency factors.

\begin{table}
\caption{\label{tabselect}
Summary of the event selection procedures
as well as their differential
suppression and efficiency factors.
}
\begin{ruledtabular}
\begin{tabular}{lcc}
Cuts & Suppression & Efficiency  \\ \hline
Raw Data  & 1.0 & 1.0 \\
Anti-Compton (AC)  & 0.063 & 0.994 \\
Cosmic-Ray Veto (CRV) & 0.957 & 0.948 \\
Pulse Shape Analysis  & 0.860 & 0.996 \\ \hline
Cumulative Net & 0.052 & 0.938 \\ 
\end{tabular}
\end{ruledtabular}
\caption{\label{tabsys}
The various contributions to the systematic
uncertainties.
}
\begin{ruledtabular}
\begin{tabular}{lcc}
Contributions & Uncertainties & $\delta ( \ke10 ^ 2 )$ \\ \hline
Relative normalization ON/OFF & $<$0.2\% & $<$0.30 \\
Reactor neutrino spectra & 24\% & 0.23 \\
SM background subtraction & 23\% & 0.03 \\ 
Efficiencies of MM signal & $<$0.2\% & $<$0.01 \\ \hline
Combined Systematic Error & $-$  & $<$0.38 \\
\end{tabular}
\end{ruledtabular}
\end{table}

\begin{figure}
\includegraphics[width=7cm]{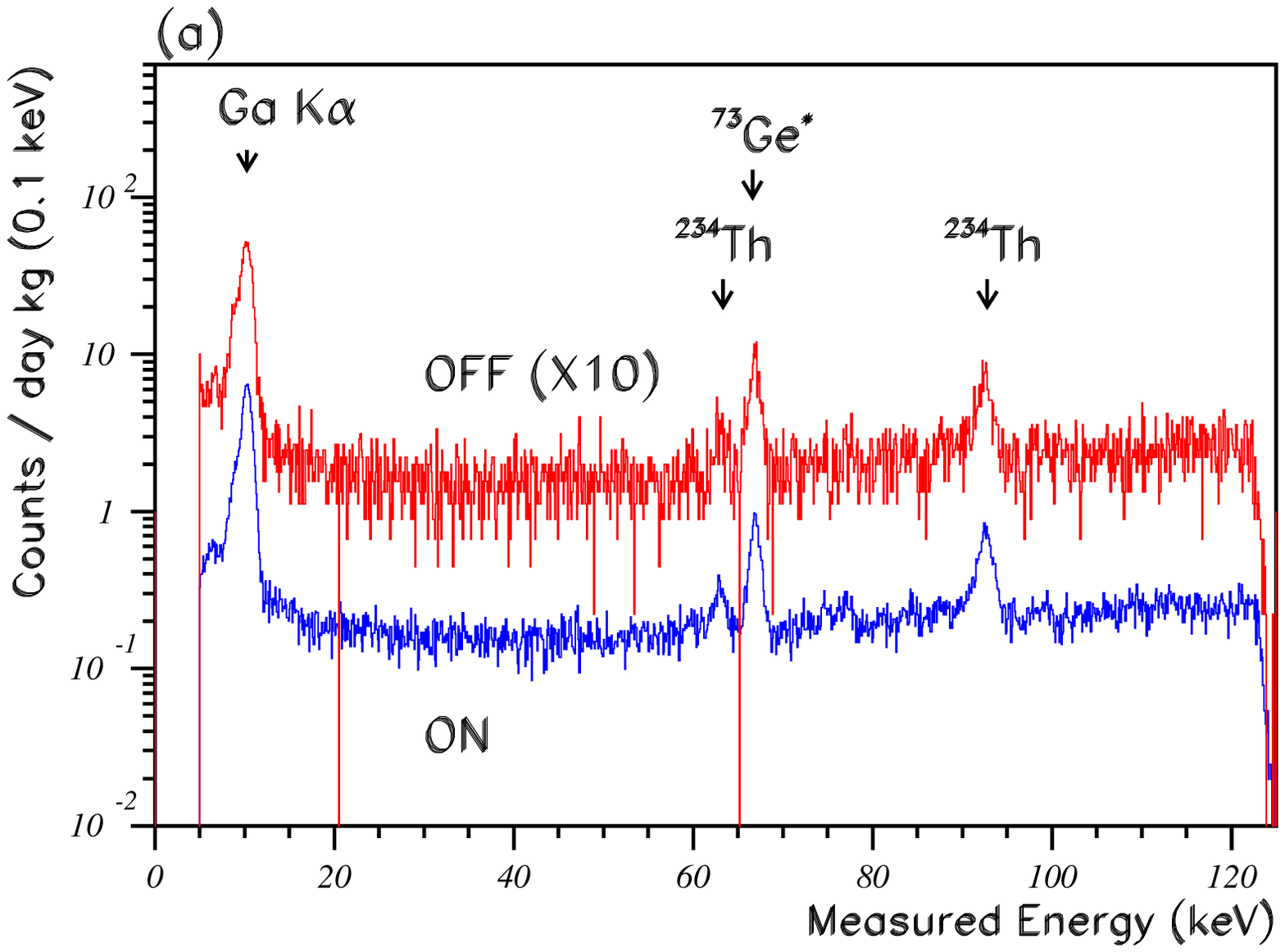}\\
\includegraphics[width=7cm]{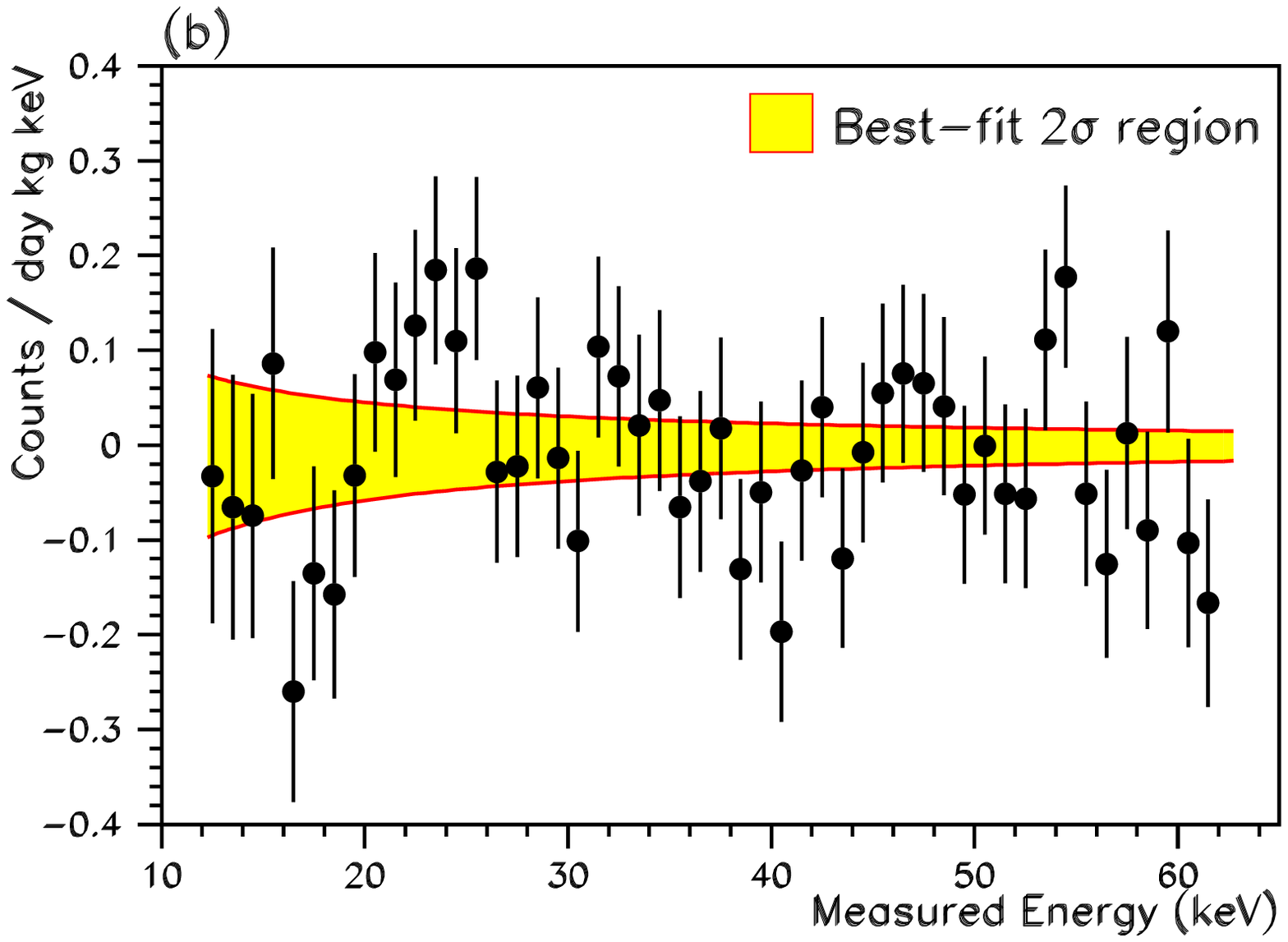}
\caption{
(a)
The energy spectra for lone events, and
(b)
The residual of the
Reactor ON recoil spectrum over
$\rm{\phi_{OFF}}$,
from 4712/1250 live time hours of 
Reactor ON/OFF data taking.
}
\label{onoff}
\end{figure}

The lone-event spectra from 
for 4712/1250 live time hours 
of reactor ON/OFF data 
are displayed in Figure~\ref{onoff}a.
The relative normalization is known to $<$0.2\%.
A detector threshold of 5~keV
and a background of
$\sim$1 keV$^{-1}$kg$^{-1}$day$^{-1}$ 
above 12~keV were achieved.
This is comparable to the typical range 
in underground Cold Dark Matter experiments.
Several lines can be identified: 
Ga X-rays at 10.37~keV and $^{73}$Ge$^*$ at 66.7~keV
from internal cosmic-induced activities, and
$^{234}$Th at 63.3~keV and 92.6~keV due to residual ambient
radioactivity from the $\u238$ series in the vicinity of
the target.
The ON and OFF spectra differ only in one significant feature.
There is a difference in
the Ga X-ray peaks, which originates
from the long-lived isotopes 
($^{68}$Ge and $^{71}$Ge with half-lives of 
271 and 11.4 days, respectively)  
activated by cosmic-rays 
prior to installation. 
The time evolution of the peak intensity can be fit to 
two exponentials consistent with the two known
half-lives.
A threshold of 12~keV was therefore adopted for
physics analysis. This is also above the energy range
where corrections due to atomic effects have
to be considered.

The reactor neutrino spectrum and its time
dependence were evaluated from
reactor operation data using the standard prescriptions 
on $\nuebar$ from fissions~\cite{vogelengel}
together with a low energy contribution due to 
neutron capture on $\u238$~\cite{rnule}.
The total flux at the detector 
is $\rm{5.8 \times 10^{12} ~ cm^{-2} s^{-1}}$,
corresponding to a MS event rate 
of 5.9~$\rm{kg^{-1} day^{-1}}$ above 12~keV at $\ke10=2$. 
The uncertainties of the $\nuebar$ spectrum at low energy are not
well studied~\cite{lernu}. We take the reasonable
ranges of 5\% and 30\% to be the systematic errors 
above and below 2~MeV, respectively.

The electron recoil spectra from SM and MS
interactions at $\munuebar = 10^{-10} ~ \mub$,
denoted by $\rm{ \phi^{SM} }$ and $\rm{\phi_{-10}^{MS} }$, respectively,
were evaluated from $\nuebar$ spectrum.
The background in the energy range of 12 to 60~keV are 
due to Compton scatterings of higher energy $\gamma$'s.
The OFF spectrum was
fitted to a 4th-degree polynomial function $\rm{\phi_{OFF}}$,
and a $\chi ^2$/dof of 80/96 was obtained.
The function $\rm{\phi_{OFF}}$ and its uncertainties
were then used as input to the fit of the ON spectrum to 
$\rm{
\phi_{OFF} +  \epsilon ( \phi^{SM} + \ke10 ^2 ~ \phi_{-10}^{MS} )
}$,
where $\epsilon$ is the analysis efficiency
of Table~\ref{tabselect}.
A best-fit value of 
$\rm{ \ke10 ^2 = -0.41 \pm  1.28 (stat.) \pm  0.38 (sys.)}$ at
$\chi ^2$/dof of 48/49  was obtained.
The various  systematic errors and their contributions
to $\delta ( \ke10 ^2 )$ are summarized in Table~\ref{tabsys}.
Following the unified approach~\cite{pdg},
the limits on the $\nuebar$ magnetic moment of
$\rm{
\munuebar  ~ < ~  1.3 (1.0) \times 10^{-10} \mub
}$
at 90(68)\% CL were derived.
The residual plot with the ON spectrum over 
$\rm{\phi_{OFF}}$ and the best-fit
2$\sigma$ region are depicted in Figure~\ref{onoff}b. 
It has been checked that positive signals
with the correct strength and uncertainties 
can be reconstructed by the procedures
from simulated spectra.

\begin{figure}
\includegraphics[width=7cm]{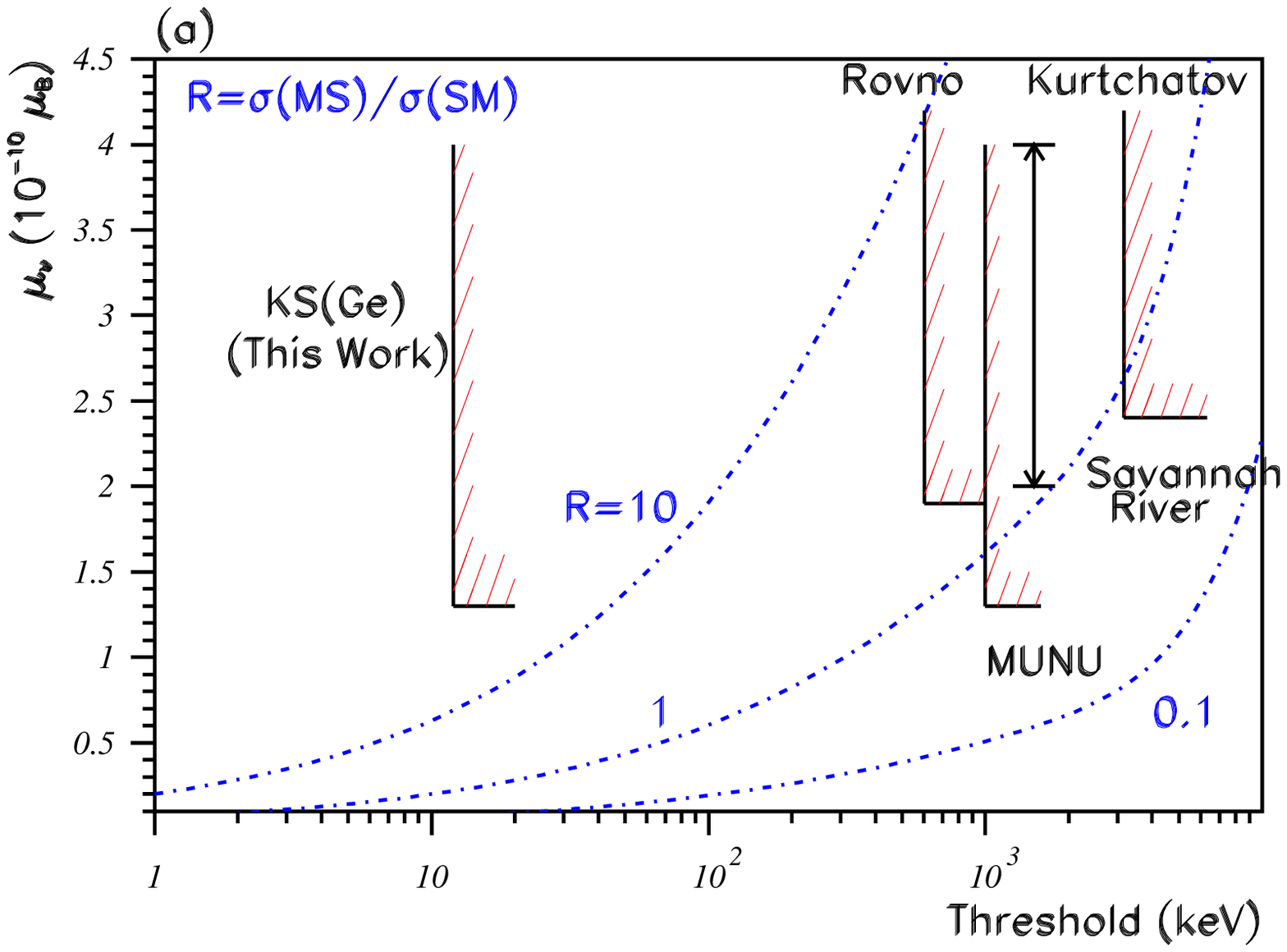}\\
\includegraphics[width=7cm]{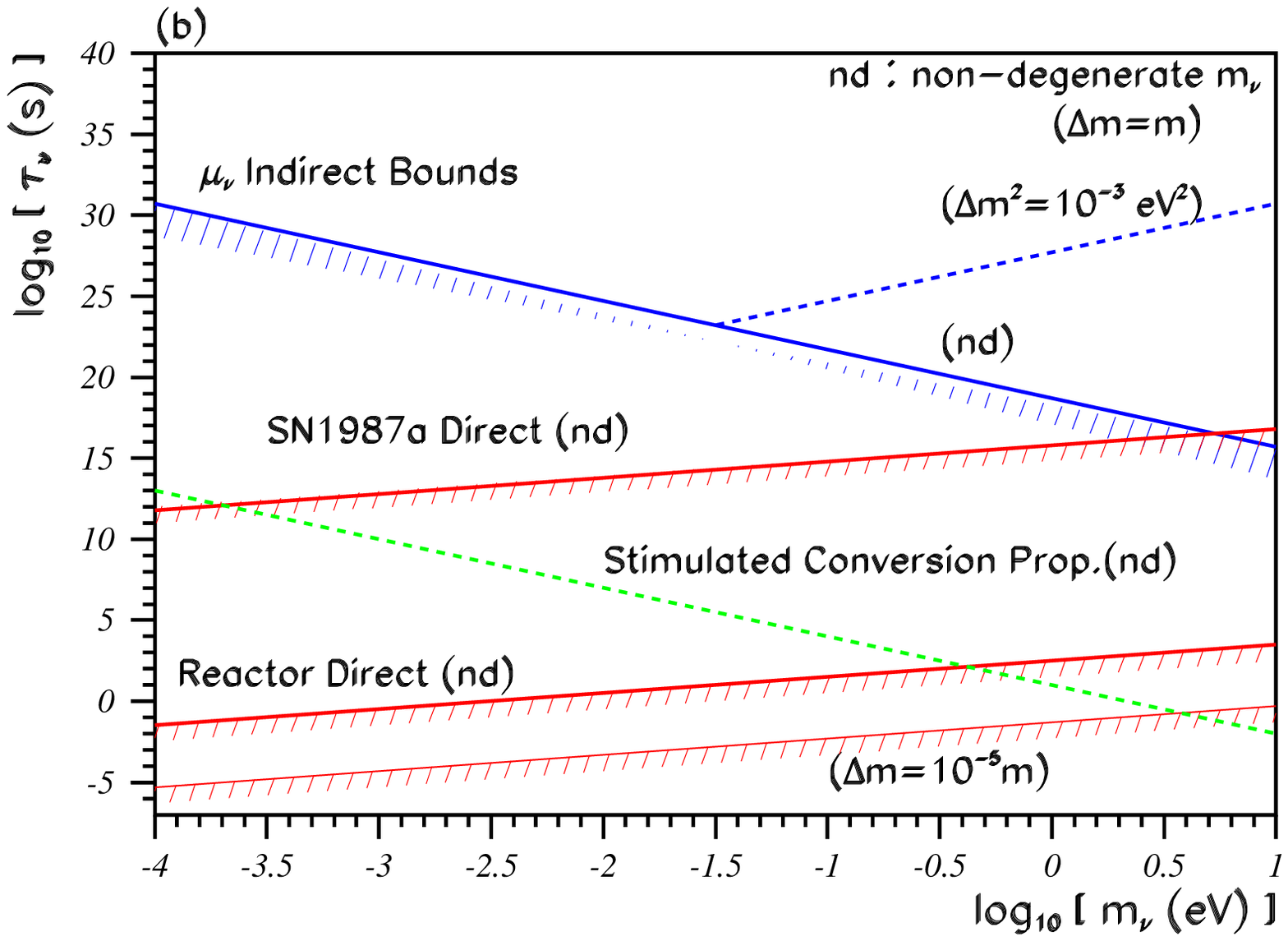}
\caption{
Summary of the results in 
(a) the searches of neutrino
magnetic moments with reactor neutrinos, 
and 
(b) the bounds of
neutrino radiative decay lifetime. 
See text for explanations.
}
\label{summaryplots}
\end{figure}

Depicted in Figure~\ref{summaryplots}a is the
summary of the results in $\rm{\munuebar}$ searches
with reactor $\nuebar$ versus the achieved
threshold. 
The dotted lines are the 
$\rm{\sigma_{MS}/\sigma_{SM}}$ ratio at a
particular (T,$\rm{\munuebar}$). 
The KS(Ge) experiment operated
at a much lower threshold of 12~keV compared
to the other efforts.
The large $\rm{\sigma_{MS}/\sigma_{SM}}$
ratio implies that the KS results
are robust against the uncertainties in the 
SM cross-sections. In particular, in
the case where the excess of events reported
in Refs.~\cite{reines} and \cite{munu} are due
to unaccounted sources of neutrinos, the
limits remain valid.

Indirect bounds on the neutrino radiative decay
lifetimes are inferred 
and displayed in Figure~\ref{summaryplots}b
for the simplified scenario where a single channel
dominates the transition, which corresponds to 
$\rm{
\tau_{\nu} m_{\nu} ^ 3 > 2.8(4.8) \times 10^{18} ~  eV ^3 s 
}$ at 90(68)\% CL in the non-degenerate case.
Superimposed are
the limits from the previous direct searches of excess
$\gamma$'s from reactor neutrinos~\cite{rdkdirect} 
and from the supernova SN1987a~\cite{rdksn}, as well
as the sensitivity level of proposed
simulated conversion experiments at accelerators~\cite{simcon}.
It can be seen that 
$\nu$-e scatterings give much more 
stringent bounds than the direct approaches.

The KS experiment continues data taking in 2002-03 using 
HPGe with improved shieldings and 186~kg
of CsI(Tl) crystals. 
Besides improving on the $\munuebar$ sensitivities, 
the goals are to perform a measurement of the
$\nuebar$-e cross-section at the MeV range, and to
study various standard and anomalous neutrino interactions.
A prototype HPGe with sub-keV threshold is being
studied.

The authors are grateful to their technical staff 
for invaluable contributions,
and to the CYGNUS Collaboration for
the veto scintillator loan.
This work was supported by contracts
90-2112-M-001-037 and 91-2112-M-001-036
from the National Science Council, Taiwan,
and 19975050 from the
National Science Foundation, China.
H.B.~Li is supported by contracts
NSC~90-2112-M002-028 and MOE~89-N-FA01-1-0
under P.W.Y.~Hwang.

\end{document}